# Thermal transports of one-dimensional ultrathin carbon structures

Jing Xue[1,2], Yuee Xie[1,2]*, Qing Peng[3] and Yuanping Chen[1,2]*

[1]*School of Physics and Optoelectronics, Xiangtan University, Xiangtan, 411105, Hunan, China*
[2] *Faculty of Science, Jiangsu University, Zhenjiang, 212013, Jiangsu, China*
[3]*Nuclear Engineering and Radiological Sciences University of Michigan, Ann Arbor, MI 48109, U.S.A.*

## Abstract

Carbon atomic chain, linear benzene polymers, and carbon nanothreads are all one-dimensional (1D) ultrathin carbon structures. They possess excellent electronic and mechanical properties; however, their thermal transport properties have been rarely explored. Here, we systematically study their thermal conductance by combining the nonequilibrium Green's function and force field methods. The thermal conductance varies from 0.24 to 1.00 nW/K at 300 K, and phonon transport in the linear benzene polymers and carbon nanothreads is strongly dependent on the connectivity styles between the benzene rings. We propose a simple 1D model, namely force-constant model, that explains the complicated transport processes in these structures. Our study not only reveals intrinsic mechanisms of phonon transport in these carbon structures, but also provides an effective method to analyze thermal properties of other 1D ultrathin structures made of only several atomic chains.

---

Corresponding authors: xieyech@xtu.edu.cn；chenyp@ujs.edu.cn.

# 1. Introduction

Graphene is an ideal electronic and thermal transport material because of its superhigh electronic and thermal conductivities [1-3]. However, a standard transport system is a one-dimensional (1D) structure [4,5], in which the carriers (electron or phonon) propagate from one side to another. Moreover, the increasing miniaturization of devices requires thin channels [6-9]. Therefore, graphene needs to be cut into 1D graphene nanoribbons in the real applications [10-14]. Figure 1(b) presents the thinnest armchair-edged nanoribbon (TAN), which seems like a string of benzene ring. The thinnest imaginable channel is a single atomic chain (SAC) [15,16], as shown in Fig. 1(a). Besides cutting graphene, another method to get ultrathin carbon channels is to construct linear benzene polymers or carbon nanothreads by linking benzene rings [17-20]. Many linearly polymerized benzene arrays and topologically distinct nanothreads have been predicted theoretically [21-23]. Some examples from two-bond linking to four-bond and six-bond linking of each benzene ring are illustrated in Figs. 1(c-h). The synthesis of these ordered nanostructures is a grand challenge task in nearly one century, because all products obtained by many ways were amorphous carbon materials [24-28]. Recently, Fitzgibbons et al. have for the first time recovered ordered products by high-pressure solid-state reaction of benzene [29], showing a promising way to fabricate all the structures in Figs. 1(c-h).

With severe quantum confinement, 1D materials, especially ultrathin nanostructures, always possess exotic physical properties and technological applications [30-38]. For example, although bulk polyethylene is a thermal insulator, the thermal conductivity of an individual polymer chain can be very high, even divergent in some cases [30-32]. The thermal conductance of gold single atom junctions is quantized at room temperature and shows that the Wiedemann-Franz law relating thermal and electrical conductance is satisfied [33]. The carbon nanothreads promise extraordinary mechanical properties such as strength and stiffness higher than that of *sp$^2$* carbon nanotubes or conventional high-strength polymers [36]. Despite the importance, the thermal transport properties of the ultrathin carbon structures in Fig. 1 have been rarely explored.

Here, we systematically study thermal conductance of ultrathin carbon structures, by combining the nonequilibrium Green's function method and force field method. Phonon dispersions of these structures are calculated firstly to confirm their structural stabilities. The calculation results also provide force constant matrices for each structure. Thermal conductance of the structures in Fig. 1 varies from 0.24 to 1.00 nW/K at 300 K, which are strongly dependent on the connectivity styles between units. The linear benzene polymer in Fig. 1(f) has the lowest thermal conductance, while the nanothread in Fig. 1(g) exhibits the highest thermal conductance. We introduce a model to explain the dependence of thermal transports on the connectivity styles of the structures, where we simplify the structures to a SAC. By analyzing the reduced force constants (RFCs) between the atoms/units, different transport processes in the structures are revealed. This provides a method to analyze phonon transport in 1D nanostructures.

## 2. Model and methods

The eight 1D ultrathin carbon structures we studied are shown in Fig. 1. Figure 1(a) is the SAC linked by carbon atoms one by one. When each atom in Fig. 1(a) is replaced by a hexagonal carbon ring, one can get a structure in Fig. 1(b), i.e., a TAN. The boundary atoms are hydrogenated for stabilization. Figures 1(c-h) present six relatively complicated structures. All the six structures can be considered as a string of benzene rings. The difference between them is the connectivity style, such as degree of saturation and topochemical pattern. We use a Roman numeral to represent degree of saturation, i.e., the number of four-coordinate carbon atoms per benzene formula unit. The structures in Figs. 1(c-d) are labeled as II, while those in Figs. 1(e-f) and 1(g-h) are labeled as IV and VI, respectively. Usually, the structures in Figs. 1(g-h) are also called carbon nanothreads because all the carbon atoms are four-coordinate, similar to the case in the diamond threads. Other degree structures like those in Figs. 1(c-f) are called linear benzene polymers. The Arabic numeral is used to represent topological pattern. We only select two representative stable structures for each type.

We use the nonequilibrium Green's function (NEGF) [39,40] to study the ballistic phonon transport properties of the 1D structures. Each structure has three parts: left lead, right lead, and the center scattering region. In order to investigate intrinsic transport properties of the structures, the three parts are set to be homogeneous to avoid scattering between them. According to the NEGF scheme, the retarded Green's function is expressed as [39]:

$$G^r = [(\omega + i0^+)^2 I - K^C - \Sigma_L^r - \Sigma_R^r]^{-1}, \qquad (1)$$

where $\omega$ is the frequency of phonons, $I$ is the identity matrix, $K^C$ is the mass-weighted harmonic force constants matrix of the center region, and $\Sigma_\beta^r = V^{C\beta} g_\beta^r V^{\beta C}$ ($\beta = L, R,$ corresponding to the left and right) denotes the self-energy of the (left and right) lead $\beta$, in which $V^{C\beta} = (V^{\beta C})^T$ is the coupling matrix of the lead $\beta$ to the center region and $g_\beta^r$ is the surface Green's function of the lead. The second-generation reactive empirical bond order potential is adopted to optimize the geometric structure and obtain the force constants of the structures. Since this potential holds long-range character, the length of the central region is set long enough to avoid interaction between the left and right thermal leads. Once the retarded Green's function $G^r$ is obtained, we can calculate the transmission coefficient $T[\omega]$ and then the thermal conductance $\kappa$ of the structures:

$$T[\omega] = Tr\{G^r \Gamma_L G^a \Gamma_R\}, \qquad (2)$$

$$\kappa(T) = \frac{\hbar}{2\pi} \int_0^\infty T[\omega]\, \omega\, \frac{\partial f(\omega)}{\partial T} d\omega, \qquad (3)$$

where $G^a = (G^r)^+$ is the advanced Green's function and $\Gamma_\beta = i(\Sigma_\beta^r - \Sigma_\beta^a)$ is the coupling function of the $\beta$ lead, and $f(\omega) = \left\{\exp\left[\frac{\hbar\omega}{k_B T}\right] - 1\right\}^{-1}$ is the Bose–Einstein distribution function for the heat carrier at the leads.

As mentioned above, the unit cells of all the structures in Fig. 1 are optimized by dynamics software "General Utility Lattice Program" (GULP) [41], based on force field method, with the second-generation reactive empirical bond order potential [42] which was proven to give excellent description of carbon-carbon bonding interactions [43-44]. The phonon dispersions are calculated to test the stabilities of these structures, and the results are shown in Fig. S1 in the supplementary materials (SM). One can find that there is no imaginary frequency appearing in the plots, which indicates that these 1D carbon structures are stable. Then, the force constants can be given by the second derivatives with respect to the potential energy. We also examine thermal stabilities of these structures by performing ab initio molecular dynamics (AIMD) simulations in a canonical ensemble[45]. After heating up to the targeted temperature of 1000 K for 20 ps, one can find that the deformations of all structures are still very small (see Fig. S4 in SM).

## 3. Results and discussion

Thermal conductance $\kappa$ of the eight ultrathin carbon structures increases with the temperature, as shown in Fig. 2, because the phonon modes are gradually excited by temperature. The thermal conductance of the SAC is 0.68 nW/K at $T = 300$ K. Although the other seven structures are all made of benzene rings, their thermal conductance has considerable difference, ranging from 0.24 to 1.00 nW/K (at T = 300 K). The thermal conductance of nanothread VI-1 is the highest, which is approximate five times of that of linear benzene polymer IV-2, implying that the degree of saturation and topological form have important effects on the phonon transport in these structures. In the following, we explain the origination of these transport differences.

The first question is how to describe thermal transport properties of these ultrathin structures. To answer this question, one can firstly consider the three simplest structures in Figs. 1(a-c). It is easy to describe the SAC by a homogeneous RFC between two neighboring carbon atoms. The purple line in Fig. 3(b) shows the value of the RFC, while the top panel in Fig. 3(a) presents the schematic model. For the TAN in Fig. 1(b), if each ring is shrunk to a lattice, it is equivalent to a SAC. The difference is each lattice has its own inner RFC, as shown in the middle panel of Fig. 3(a). The RFCs on/between the lattices are calculated approximately according to average values of the force constant matrix in *xx*, *yy* and *zz* directions[47], i.e., only one phonon mode is need to be considered here (two examples of calculating RFC are given in Fig. S2, Tables S1 and S2 in SM). The linear benzene polymer II-1 in Fig. 1(c) is somewhat similar to the TAN in Fig. 1(b), and the only difference is the benzene rings in the former are tilted because of the existence of the saturated hydrogen atoms. Therefore, one can also use the similar simplified model to simulate phonon transport in it. The simplified model is shown in the bottom panel in Fig. 3(a), where the RFC between the lattices is smaller than that in the TAN because of the tilted benzene rings. According to the parameters of the RFC shown in Fig. 3(b), thermal conductance of these simplified models is calculated, and the results are given in Fig. 3(c), agreeing well with that in Fig. 2(b). (It is noted that the thermal conductance of models is smaller than that of the real structures, which is because in the models only one phonon mode is considered.) This indicates that one can use simplified model like atomic chains to simulate phonon transport in these ultrathin structures. It is noted that single polyethylene chain (SPC) is another simple carbon atomic chain passivated by hydrogen atoms. We compare thermal conductance of SAC with that of SPC (see Fig. S3 in SM). One can find that the former is larger than the latter, because there is no phonon scattering induced by hydrogen atoms in SAC. We also compare thermal conductance of SPC of our results at 300 K and that calculated by molecular dynamics (MD) in reference [31] (see the red dot in Fig. S3(d)).

Although the calculated methods are different, the calculated values are comparable (0.32 nW/K (our result) and 0.50 nW/K (MD)). It indicates that our method is valid.

To explain the effect of degree of saturation on thermal transport, we compare structures of II-1, IV-1, VI-1 and their thermal conductance (see Fig. 2(c)). The degrees of saturation for II-1, IV-1, VI-1 are 2, 4 and 6, respectively. Their thermal conductance is about 0.40, 0.70 and 1.00 nW/K at temperature $T = 300$ K, respectively, i.e., the latter two are approximate two and three times of the former one. We also apply the simplified force-constant models in Fig. 4(a) to simulate these 1D structures, and the calculated thermal conductance is shown in Fig. 4(c). Obviously, these thermal conductance shows substantial deviation from the real thermal conductance in Fig. 2(c). The discrepancy is attributed to different number of phonon transport channels in the structures. In the structure II-1, the two saturated bonds between benzene rings form one transport channel along $z$ direction. In the structures IV-1 and VI-1, the saturated bonds form two and three channels, respectively. Therefore, the total thermal conductance for the three structures should be 1, 2 and 3 times of those for the corresponding simplified models. After considering the channel numbers, the thermal conductance of models fits the results of real structures well (see Fig. 4(d)). By analyzing thermal transports of the three structures, the structures with multi-degree of saturation can still be simplified to a 1D chain, but multiple channels should be considered.

Finally, we discuss the effect of topological pattern on the thermal transport in the ultrathin carbon structures. We use the two structures II-1 and II-2 as examples.

Structures II-1 and II-2 have the same degree of saturation, but they have different connectivity styles between the benzene rings. They nearly have the same RFCs on the lattices as well as between the lattices, as shown the black line labeled with triangles in Fig. 5(a). In this case, one cannot use one-phonon mode to compare their thermal properties. Because the benzene rings in the two structures have different tilted directions and angles, the RFCs on/between the lattices of the simplified models have different projected weights along *xx*, *yy* and *zz* directions, as shown the colored labels in Fig. 5(a). This results in the two structures having different components of thermal conductance (see Fig.5(b)). Because the *xx* and *yy* components of structure II-1 contribute larger thermal conductance than those of structure II-2, the total thermal conductance of the former is greater than that of the latter, as shown in Fig.5(c).

## 4. Conclusions

In conclusion, we systematically study thermal conductance of 1D ultrathin carbon structures, including carbon atomic chain, linear benzene polymers, and carbon nanothreads. Their thermal conductance varies from 0.24 to 1.00 nW/K at 300 K. The phonon transports in the linear benzene polymers and carbon nanothreads are sensitive to the connectivity styles between the benzene rings, such as degree of saturation and topochemical pattern. We introduce a simple force-constant model to analyze the complicate transport processes in these structures. Our study on one hand provides basic data to evaluate transport abilities of these 1D carbon structures, on the other hand proposes an insight to understand phonon transport in 1D ultrathin nanostructures.

In this work, we only focus on thermal transport in infinite-length 1D ultrathin carbon nanostructures. The previous studies indicate that thermal transports in finite-length 1D nanostructures [48,49] and bulk van der Waals nanowires [50] possess different transport phenomena with those in infinite-length 1D nanostructures. Thus, based on our results, it is worth studying length-dependent thermal transport in finite-length 1D nanostructures and transport issues in bulk crystal made of van der Waals carbon nanowires in the future.

## Acknowledgments

This work was supported by the National Natural Science Foundation of China (No.11874314).

**Figure Captions**

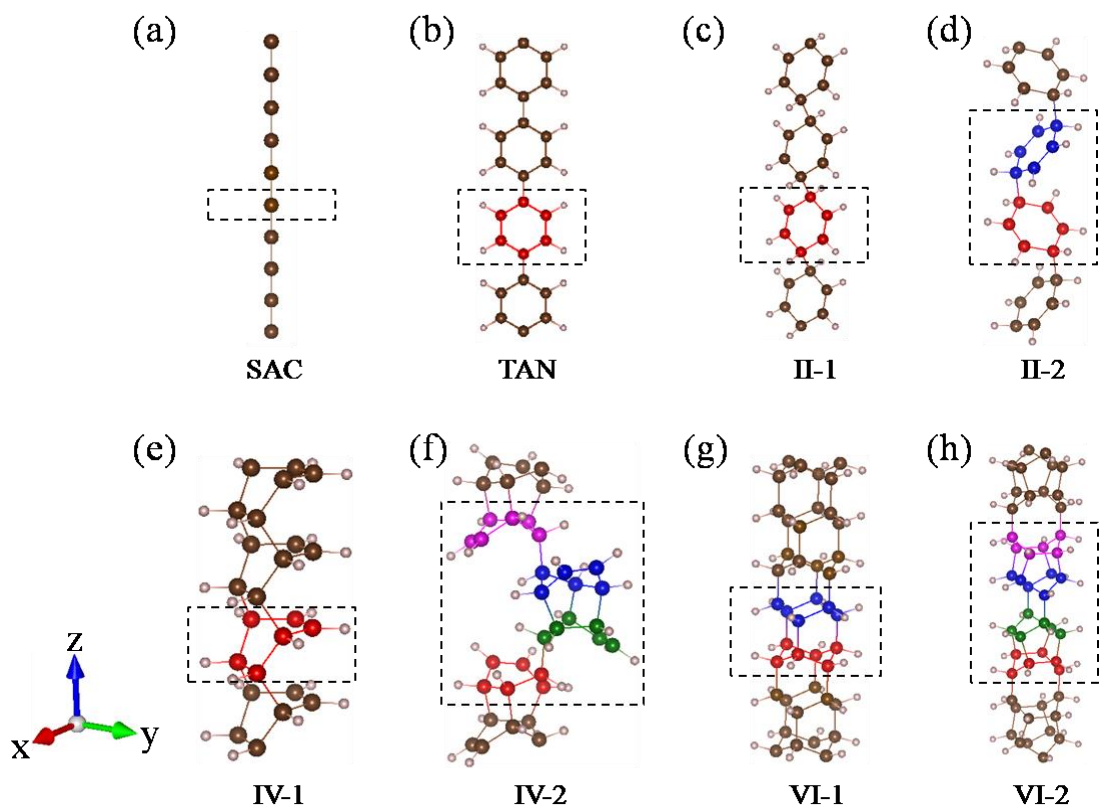

Figure 1. Atomic structures of (a) a SAC (single atomic chain), (b) a TAN (thinnest armchair-edged nanoribbon), (c-d) two linear benzene polymers II-1 and II-2, whose degree of saturation is two, (e-f) two linear benzene polymers IV-1 and IV-2, whose degree of saturation is four, (g-h) two carbon nanothreads VI-1 and VI-2, whose degree of saturation is six. The Roman numerals represent degree of saturation, while the Arabic numerals represent topological pattern. The dashed lines show primitive cells of the structures, in which the atoms with the same color represent atoms in a benzene ring.

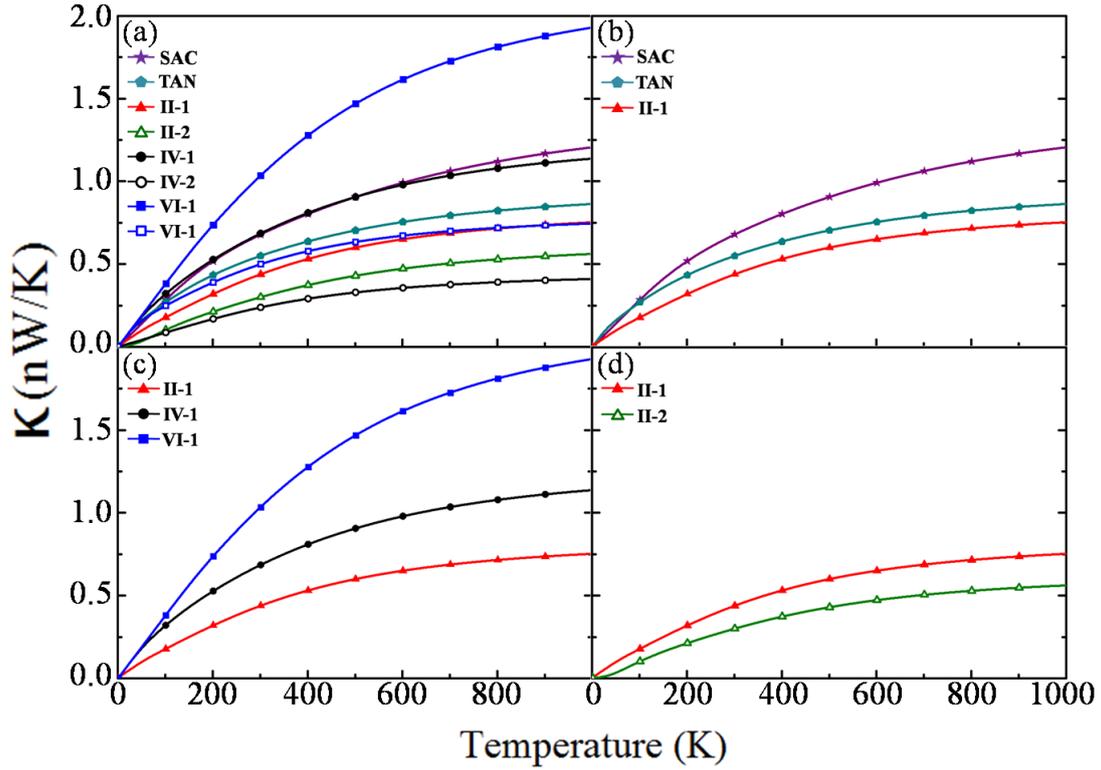

Figure 2. (a) Thermal conductance $\kappa$ of all the eight structures in Fig. 1. (b-d) Comparison of thermal conductance $\kappa$ between (b) the simplest three structures, SAC, TAN and II-1; (c) three structures II-1, IV-1 and VI-1 with different number of transport channels; (d) two structures II-1 and II-2, with the same degree of saturation but different chemical pattern.

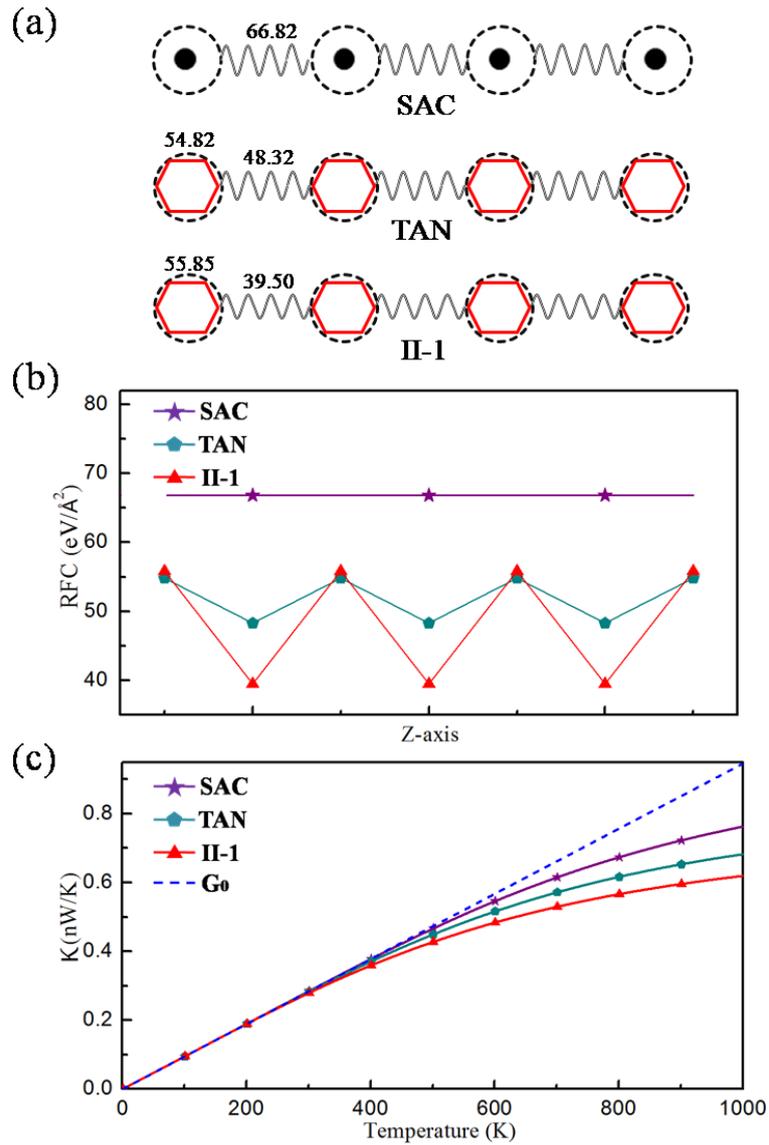

Figure 3. (a) Simplified models of the three simplest structures SAC, TAN and II-1, where the benzene rings are reduced to lattices in a chain. (b) Distributions of RFCs for the three simplified models in (a). The RFCs are also labeled in (a). (c) Thermal conductance $\kappa$ of the three structures calculated based on the simplified models in (a) and RFCs in (b). $G_0$ is the quantized thermal conductance [45], which corresponds to a biggest ideal thermal conductance of one-phonon mode.

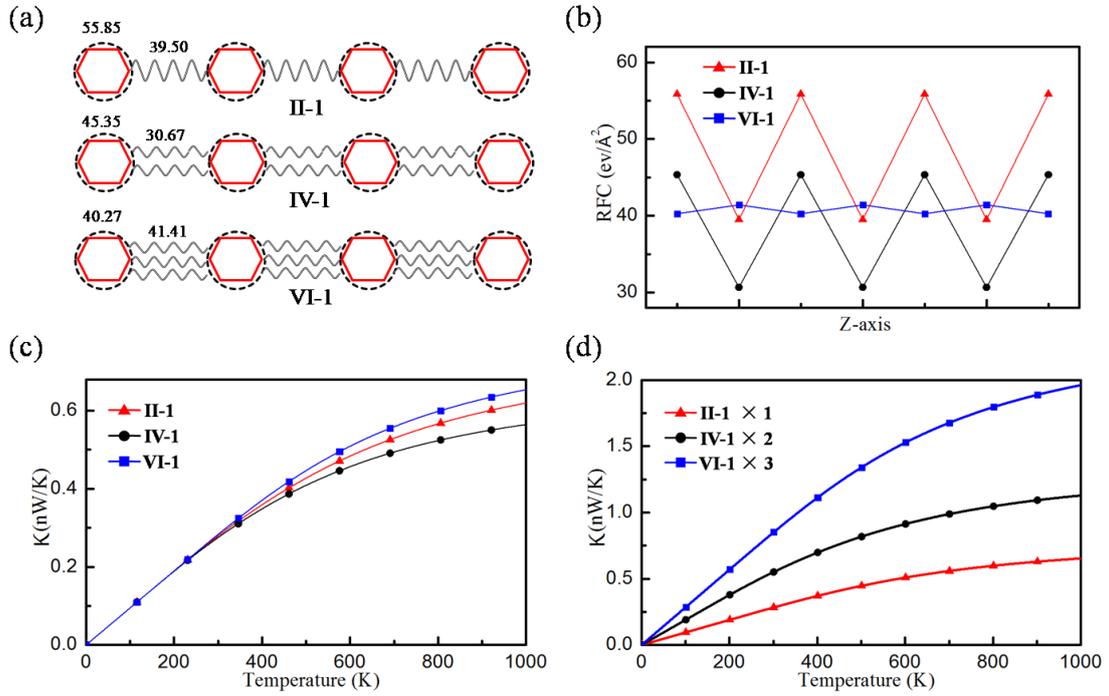

Figure 4. (a) Simplified models for the structures II-1, IV-1 and VI-1, where the number of "springs" between the lattices represent the number of channels. (b) Distributions of RFCs for the three structures. (c) Thermal conductance $\kappa$ of three simplified models with only one channel. (d) Thermal conductance $\kappa$ of the three simplified models with realistic number of channels.

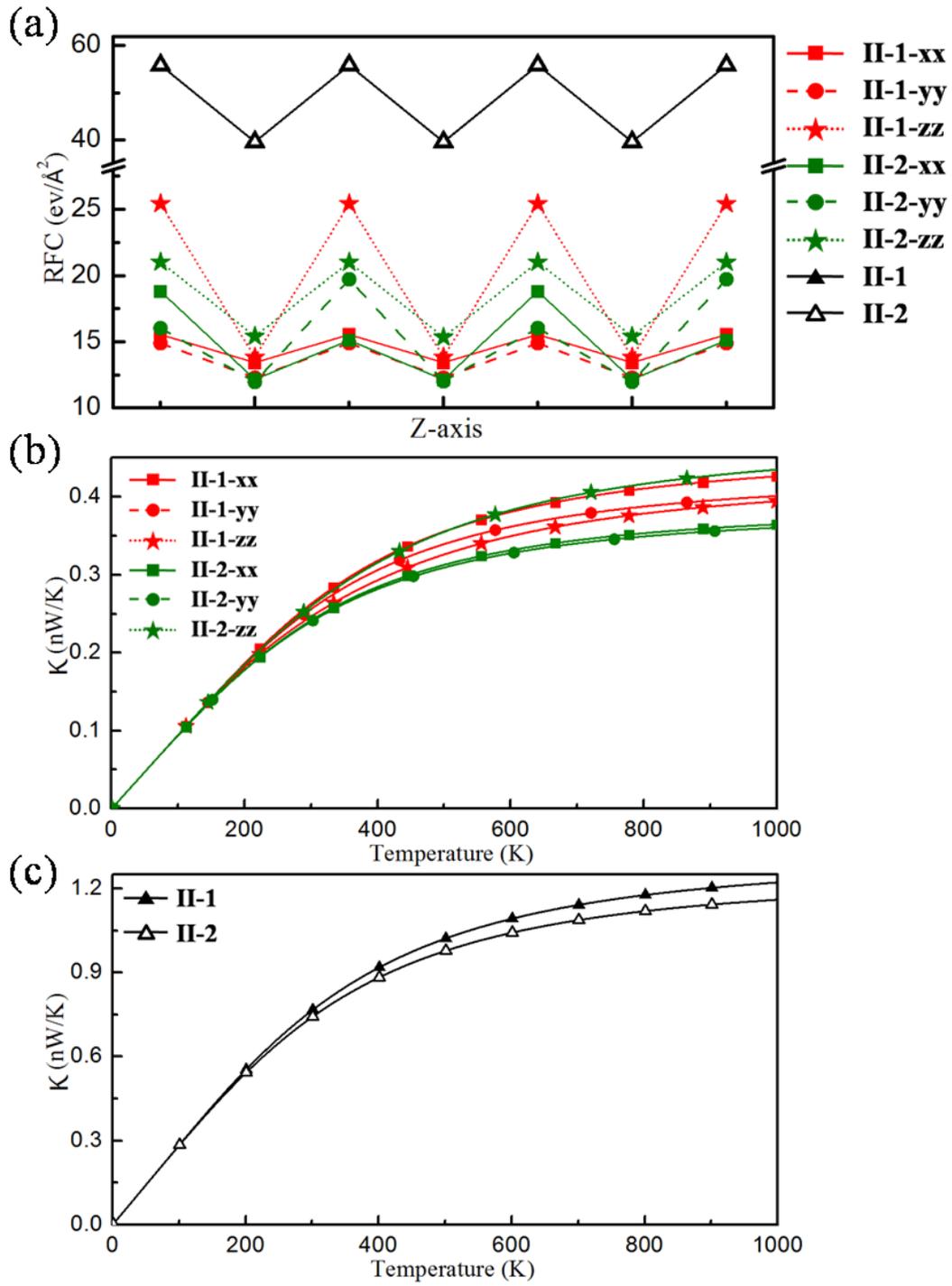

Figure 5. (a) Distribution of RFC components for the two structures II-1 and II-2. The total RFCs of the two structures are also shown as triangles for comparison. (b) Thermal conductance $\kappa$ of the two structures based on different components of RFCs in (a). (c) The total thermal conductance $\kappa$ of the two structures.